\documentclass[12pt,a4paper]{article}
\usepackage[utf8]{inputenc}
\usepackage[english]{babel}
\usepackage[top=2.5cm, bottom=3.0cm, left=2.5cm, right=2.5cm]{geometry}
\usepackage{setspace}
\makeatletter
\g@addto@macro\bfseries{\boldmath}
\makeatother
\usepackage{fancyhdr}
\usepackage{amssymb}
\usepackage{amsmath}
\usepackage{amsfonts}
\usepackage{mathrsfs}
\usepackage{bm}
\usepackage{slashed}
\usepackage{mathtools}
\usepackage{amsthm}

\numberwithin{equation}{section}
\allowdisplaybreaks
\usepackage{braket}
\usepackage{enumitem}
\usepackage{array}

\usepackage[]{caption} 
\usepackage{hyperref}
\hypersetup{
    colorlinks,%
    citecolor=blue,%
    filecolor=blue,%
    linkcolor=blue,%
   	urlcolor=black,
   	linktoc=page
}
\usepackage{cite}
\usepackage[toc,page]{appendix}
\usepackage{graphicx,datetime}
\usepackage{float}
\usepackage{subfig}
\usepackage{tikz}
\usetikzlibrary{calc}
\usetikzlibrary{decorations.pathreplacing}
\usetikzlibrary{decorations.pathmorphing}
\usetikzlibrary{decorations.markings}
\usetikzlibrary{arrows.meta}
\usetikzlibrary{positioning}
\tikzset{every picture/.style={font issue=\footnotesize},
         font issue/.style={execute at begin picture={#1\selectfont}}
        }
\usepackage{xcolor}
\definecolor{DarkGreen}{RGB}{0,128,0}
\usepackage{mdframed}

\newcommand{\poubelle}[1]{}

\renewcommand{\textbf}[1]{\begingroup\bfseries\mathversion{bold}#1\endgroup}
\newcommand{\D}{d}

 \newcommand{\badat}{\begin{alignedat}}
 \newcommand{\eadat}{\end{alignedat}}
 \def\be{\begin{equation}}
\def\ee{\end{equation}}
\newcommand{\mO}{\mathcal{O}}
\newcommand{\mW}{\mathcal{W}}
\newcommand{\mT}{\mathcal{T}}
\newcommand{\mY}{\mathcal{Y}}
\newcommand{\mNzero}{\mathcal{N}^{(0)}}
\newcommand{\mNone}{\mathcal{N}^{(1)}}
\newcommand{\bz}{\bar{z}}
\newcommand{\out}{\text{out}}
\newcommand{\outstate}{\langle \text{out}|}
\newcommand{\instate}{|\text{in}\rangle}
\newcommand{\scri}{\mathscr{I}}
\usepackage{pict2e}
\makeatletter
\newcommand{\loplus}{\mathbin{\mathpalette\dog@lsemi{+}}}
\newcommand{\dog@lsemi}[2]{\dog@semi{#1}{#2}{270,90}}
\newcommand{\dog@semi}[3]{%
  \begingroup
  \sbox\z@{$\m@th#1#2$}%
  \setlength{\unitlength}{\dimexpr\ht\z@+\dp\z@\relax}%
  \makebox[\wd\z@]{\raisebox{-\dp\z@}{%
    \begin{picture}(1,1)
    \linethickness{\variable@rule{#1}}
    \roundcap
    \put(0.5,0.5){\makebox(0,0){\raisebox{\dp\z@}{$\m@th#1#2$}}}
    \put(0.5,0.5){\arc[#3]{0.5}}
    \end{picture}%
  }}%
  \endgroup
}
\newcommand{\variable@rule}[1]{%
  \fontdimen8  
  \ifx#1\displaystyle\textfont3\else
    \ifx#1\textstyle\textfont3\else
      \ifx#1\scriptstyle\scriptfont3\else
        \scriptscriptfont3\relax
  \fi\fi\fi
}
\makeatother

\begin{document}

\setstretch{1.1}

\setcounter{tocdepth}{2}

\begin{titlepage}

\begin{flushright}
\end{flushright}
\vspace{0.5cm}


\begin{center}
\setstretch{2.2}
{\LARGE{\textbf{Loop-corrected subleading soft theorem\\ and the celestial stress tensor}}}
\end{center}


 \vspace{7mm}
\begin{center} 
Laura Donnay$^\star$\footnote{\fontsize{8pt}{10pt}\selectfont \ \href{mailto:laura.donnay@tuwien.ac.at}{laura.donnay@tuwien.ac.at}}, Kevin Nguyen$^\dagger$\footnote{\fontsize{8pt}{10pt}\selectfont\  \href{mailto:kevin.nguyen@kcl.ac.uk}{kevin.nguyen@kcl.ac.uk}}, Romain Ruzziconi$^\star$\footnote{\fontsize{8pt}{10pt}\selectfont\ \href{mailto:romain.ruzziconi@tuwien.ac.at}{romain.ruzziconi@tuwien.ac.at}}

\vspace{2mm}
\normalsize
\bigskip\medskip
$^\star$\textit{Institute for Theoretical Physics, TU Wien,\\
Wiedner Hauptstrasse 8, A-1040 Vienna, Austria\\
\vspace{2mm}
}

\normalsize
\medskip
$^\dagger$\textit{Department of Mathematics, King's College London,\\
The Strand, London WC2R 2LS, UK\\
\vspace{2mm}
}
 
 \end{center}

\begin{center}

\vspace{20pt}

\begin{abstract}\noindent
We demonstrate that the one-loop exact subleading soft graviton theorem automatically follows from conservation of the BMS charges, provided that the hard and soft fluxes separately represent the extended BMS algebra at null infinity. This confirms that superrotations are genuine symmetries of the gravitational $\mathcal{S}$-matrix beyond the semiclassical regime. In contrast with a previous proposal, the celestial stress tensor accounting for the one-loop corrections follows from the gravitational phase space analysis and does not require the addition of divergent counterterms. Moreover, we show that the symplectic form on the radiative phase space factorises into hard and soft sectors, and that the resulting canonical generators precisely coincide with the correct BMS fluxes.
\end{abstract}

\end{center}

\end{titlepage}

\newpage

\setcounter{page}{2}

\begin{spacing}{0.8}


\end{spacing}

\newpage

\tableofcontents

\section{Introduction}

Important progress has been achieved in the last decade in unraveling the infrared structure of quantum gravity (see \cite{Strominger:2017zoo,McLoughlin:2022ljp} for reviews). One of the key features is the connection between asymptotic symmetries and soft graviton theorems. More precisely, conservation of supertranslation charge was shown to be equivalent to Weinberg's leading soft graviton theorem \cite{Strominger:2013jfa,He:2014laa}, while conservation of superrotation charge was shown to encode the subleading soft graviton theorem \cite{Barnich:2009se,Cachazo:2014fwa,Kapec:2014opa}. In the context of celestial holography as a recent proposal for flat space holography (see e.g. \cite{deBoer:2003vf,Barnich:2010eb,Pasterski:2016qvg} for pioneering works and \cite{Pasterski:2021rjz,Raclariu:2021zjz} for reviews), soft theorems have been recast as Ward identities of a two-dimensional \textit{celestial} CFT. As a particular case of interest to the present work, the tree-level subleading soft graviton theorem was shown to reproduce the conformal Ward identities of a celestial stress tensor \cite{Kapec:2016jld}.

While the leading soft theorem is tree-level exact, the subleading soft theorem is one-loop exact \cite{Bern:2014oka}. The one-loop corrections to the subleading soft factor involve an infrared divergent piece, as well as a possible finite contribution. Due to the additional $\log \omega$ corrections appearing at one loop \cite{Sahoo:2018lxl,Laddha:2018vbn,Campiglia:2019wxe}, the finite contribution is scheme-dependent and only the infrared divergent piece is unambiguous. 
While the tree-level subleading soft graviton theorem has been successfully related to the conformal Ward identity of a celestial stress tensor \cite{Kapec:2014opa,Kapec:2016jld}, the inclusion of the one-loop corrections into this correspondence has not been straightforward.
This raised the question of whether superrotations are only effective symmetries of semiclassical gravity or whether they hold at the quantum level. He, Kapec, Raclariu and Strominger however suggested that the one-loop correction to the subleading soft factor could be accounted for at the cost of introducing IR-divergent one-loop correction to the celestial stress tensor itself \cite{He:2017fsb}. They were able to show that the loop-corrected subleading soft graviton theorem is then equivalent to the conformal Ward identity of the corrected stress tensor, henceforth restoring the Virasoro superrotation symmetries at the quantum level.

Since this pioneering analysis, a lot of progress has been made in understanding the gravitational phase space in presence of superrotations. This includes the renormalisation of the symplectic structure at $\mathscr{I}^+$ \cite{Compere:2018ylh}, the identification of the soft modes in the phase space with the superboost field \cite{Compere:2016jwb,Compere:2018ylh}, a refinement of the radiative phase space analysis \cite{Alessio:2019cae, Campiglia:2021bap}, and the improvement of the BMS fluxes \cite{Campiglia:2020qvc,Compere:2020lrt,Donnay:2021wrk, Fiorucci:2021pha}. In particular, the improved BMS fluxes were shown to satisfy the following properties: $(i)$ they form a representation of the BMS algebra for the standard bracket, $(ii)$ they vanish in absence of radiation, $(iii)$ they are finite provided one appropriately restricts the falloffs in the limit to the corners of $\scri^+$. 

A new split between hard and soft BMS momentum fluxes was prescribed in \cite{Donnay:2021wrk} so that the hard and soft sectors transform separately in the coadjoint representation of the extended BMS algebra \cite{Barnich:2021dta}. The corrections with respect to the standard fluxes involve two key features: $(i)$ the covariance with respect to the action of superrotation symmetries with a proper treatment of the superboost field, and $(ii)$ the addition of soft terms which allow to reproduce the correct transformation laws under the action of supertranslations. The main result of this work is that the corrections proposed in \cite{Donnay:2021wrk} for the hard and soft fluxes automatically account for the one-loop correction to the subleading soft factor, without the need to introduce any explicit IR divergent terms in contrast to the original proposal made in \cite{He:2017fsb}. The possibility of accounting for one-loop corrections in this way has also been suggested in \cite{Pasterski:2022jzc}.  

The paper is organized as follows. In Section \ref{sec:Gravity in $4d$ asymptotically flat spacetimes}, we review the phase space of gravity in asymptotically flat spacetimes. In particular, we provide an updated account on the appropriate prescription for the BMS charges and relate these expressions with the components of the Weyl tensor in Newman-Penrose formalism. This allows us to match the prescriptions of \cite{Compere:2018ylh,Campiglia:2020qvc,Ruzziconi:2020cjt,Compere:2020lrt,Donnay:2021wrk, Fiorucci:2021pha} with the one considered in \cite{Freidel:2021qpz,Freidel:2021dfs,Freidel:2021ytz}. In Section \ref{sec:Radiative phase space with superrotations}, we define the variables of the radiative phase space which allow to write the Campiglia-Laddha symplectic structure in Darboux form. We show that the latter factorises between hard and soft sectors and differs from the Ashtekar-Streubel expression by a corner ambiguity. We then show that the resulting canonical generators of BMS symmetries precisely correspond to the hard and soft BMS fluxes considered in \cite{Donnay:2021wrk}, which constitutes a key result in our analysis. In Section \ref{sec:Leading soft theorem and soft factorisation}, we review the relation between supertranslation symmetries and leading soft theorem, as well as the soft factorisation of the $\mathcal{S}$-matrix. In Section \ref{sec:Subleading soft theorem}, we rederive the relation between superrotation Ward identities and subleading soft theorems with the corrected hard and soft BMS fluxes of \cite{Donnay:2021wrk}. We demonstrate that the one-loop corrections to the subleading soft factor come out automatically, and thus promote superrotations to symmetries of quantum gravity. In Section \ref{sec:Celestial stress-tensor}, we reformulate the subleading soft theorem as a conformal Ward identity by introducing the two-dimensional celestial stress tensor which accounts for loop corrections. 
Appendix \ref{sec:Newman-Penrose formalism vs Metric formalism} provides a dictionary between metric and Newman-Penrose quantities. Appendix \ref{app:massive} contains some details regarding the one-loop corrections to the subleading soft factors when external particles are massive.

\section{Gravity in asymptotically flat spacetimes}
\label{sec:Gravity in $4d$ asymptotically flat spacetimes}

We start with a review of four-dimensional asymptotically flat spacetimes in Bondi gauge. We follow the notations and conventions of \cite{Donnay:2021wrk}. 

In Bondi coordinates $(u, r, x^A)$, $x^A = (z, \bar{z})$ \cite{Bondi:1962px,Sachs:1962wk,Sachs:1962zza}, the large-$r$ expansion of four-dimensional asymptotically flat metrics reads
\begin{equation}
   \begin{split}
        ds^2 = \, &\left(\frac{2M}{r}+\mathcal O(r^{-2})\right) \D u^2 - 2 \left( 1+\mathcal O(r^{-2})\right) \D u \D r\\
        &+ \left(r^2 \mathring{q}_{AB} + r\, C_{AB} + \mathcal{O}(r^{0})\right) \D x^A \D x^B  \\
    &+ \left(\frac{1}{2}D_B C^B_A + \frac{2}{3r}(N_A + \frac{1}{4}C_A^B D_C C^C_B) + \mathcal O(r^{-2}) \right)\D u \D x^A \, ,
    \end{split}
    \label{Bondi gauge metric}
\end{equation} where for simplicity we chose the representative of the conformal boundary metric to be the flat metric
\begin{equation}
\mathring{q}_{AB}\, dx^A dx^B = 2 dz d\bar{z}\,.
\label{sphere metric boundary}
\end{equation} Indices $A, B, C$ are lowered and raised with \eqref{sphere metric boundary} and its inverse. In these coordinates the Minkowski metric is $ds^2 = -2 du dr + 2 r^2 dz d\bar{z}$. The subleading terms in \eqref{Bondi gauge metric} with respect to the flat metric involve the Bondi mass aspect $M(u,z,\bar z)$, the angular momentum aspect $N_A (u,z,\bar z)$ and the asymptotic shear $C_{AB}(u,z,\bar z)$ which is symmetric and trace-free. The Bondi mass and angular momentum aspects satisfy the retarded time evolution/constraint equations
\begin{equation}
\begin{split}
\partial_u M &= - \frac{1}{8} N_{AB} N^{AB} + \frac{1}{4} \partial_A \partial_B N^{AB} \, , \\ 
\partial_u N_A &= \partial_A M + \frac{1}{16} \partial_A (N_{BC} C^{BC}) - \frac{1}{4} N^{BC} \partial_A C_{BC} \\
&\quad  -\frac{1}{4} \partial_B (C^{BC} N_{AC} - N^{BC} C_{AC}) - \frac{1}{4} \partial_B \partial^B \partial^C C_{AC}+ \frac{1}{4} \partial_B \partial_A \partial_C C^{BC} \, ,
\end{split}\label{EOM1} 
\end{equation} 
with $N_{AB} = \partial_u C_{AB}$ the Bondi News tensor characterising outgoing gravitational radiation. 

The diffeomorphisms preserving the falloff behavior of \eqref{Bondi gauge metric} are generated by vectors fields $\xi = \xi^u \partial_u + \xi^z \partial + \xi^{\bar{z}} \bar{\partial}+ \xi^r \partial_r$ whose leading order components are
\begin{equation}
\begin{split}
    &\xi^u =\mathcal{T} + \frac{u}{2} (\partial \mathcal{Y} +  \bar{\partial} \bar{\mathcal{Y}}) \, , \\
    &\xi^z = \mathcal{Y} + \mathcal{O}(r^{-1})\, , \qquad \xi^{\bar{z}} = \bar{\mathcal{Y}} + \mathcal{O}(r^{-1}) \, , \\ 
    &\xi^r = - \frac{r}{2} (\partial \mathcal{Y} +  \bar{\partial} \bar{\mathcal{Y}}) + \mathcal{O}(r^0) \, ,
    \end{split}
    \label{AKV Bondi}
\end{equation} where $\mathcal{T}= \mathcal{T}(z, \bar{z})$ is the supertranslation parameter and $\mathcal{Y} = \mathcal{Y}(z)$, $\bar{\mathcal{Y}} = \bar{\mathcal{Y}}(\bar{z})$ are the superrotation parameters satisfying the conformal Killing equation $\bar{\partial} \mathcal{Y} = 0= \partial \bar{\mathcal{Y}}$.  Using the modified Lie bracket $[\xi_1, \xi_2]_\star =[\xi_1, \xi_2] - \delta_{\xi_1}\xi_2 + \delta_{\xi_2} \xi_1$ where the last two terms take into account the field-dependence of the asymptotic Killing vectors \eqref{AKV Bondi} at subleading orders in $r$ \cite{Barnich:2010eb}, the asymptotic Killing vectors \eqref{AKV Bondi} satisfy the commutation relations
\begin{equation}
    [\xi (\mathcal{T}_1, \mathcal{Y}_1, \bar{\mathcal{Y}}_1), \xi (\mathcal{T}_2, \mathcal{Y}_2, \bar{\mathcal{Y}}_2)]_\star = \xi (\mathcal{T}_{12}, \mathcal{Y}_{12}, \bar{\mathcal{Y}}_{12}) \, ,
    \label{commutation relations 1}
\end{equation} with
\begin{equation}
\begin{split}
    \mathcal{T}_{12} &= \mathcal{Y}_1 \partial \mathcal{T}_2 - \frac{1}{2} \partial \mathcal{Y}_1 \mathcal{T}_2  - (1 \leftrightarrow 2) + c.c.  \, , \\
    \mathcal{Y}_{12} &= \mathcal{Y}_1 \partial \mathcal{Y}_2 - (1 \leftrightarrow 2)\,, \quad \bar{\mathcal{Y}}_{12} = \bar{\mathcal{Y}}_1 \bar{\partial} \bar{\mathcal{Y}}_2 - (1 \leftrightarrow 2) \, ,
\end{split}
    \label{commutation relations 2}
\end{equation} where ``$c.c.$'' stands for complex conjugate terms. This is the extended BMS algebra \cite{Barnich:2009se, Barnich:2010eb, Barnich:2011ct}. Under Lie derivation of the metric \eqref{Bondi gauge metric} with respect to the BMS symmetries \eqref{AKV Bondi}, the radiative data $\{C_{AB},N_{AB}\}$ transform as
\begin{equation}
    \begin{split}
        &\delta_{(\mathcal{T}, \mathcal{Y}, \bar{\mathcal{Y}})} {C}_{zz} =  \left(\mathcal{Y} \partial + \bar{\mathcal{Y}} \bar{\partial} + \frac{3}{2} \partial \mathcal{Y} - \frac{1}{2} \bar{\partial} \bar{\mathcal{Y}} \right) {C}_{zz} \\
        &\qquad\qquad\qquad+\left(\mathcal{T} + \frac{u}{2} (\partial \mathcal{Y} + \bar{\partial} \bar{\mathcal{Y}}) \right) {N}_{zz} - 2  \partial^2 \mathcal{T}  - u \partial^3 \mathcal{Y}\, , \\
        &\delta_{(\mathcal{T}, \mathcal{Y}, \bar{\mathcal{Y}})} {N}_{zz} =  \left(\mathcal{Y} \partial + \bar{\mathcal{Y}} \bar{\partial} + 2 \partial \mathcal{Y} \right) {N}_{zz} + \left(\mathcal{T} + \frac{u}{2} (\partial \mathcal{Y} + \bar{\partial} \bar{\mathcal{Y}}) \right) \partial_u {N}_{zz} - \partial^3 \mathcal{Y}\,,
    \end{split} \label{transfo CN}
\end{equation} together with the complex conjugate relations for $C_{\bar{z}\bar{z}}$ and $N_{\bar{z}\bar{z}}$. From now on, we will use the holomorphic and antiholomorphic notations with respect to $x^A =(z, \bar{z})$ to write the various expressions. The complex conjugate expressions will always be implicitly assumed. 

In addition to the falloffs in $r$ provided in \eqref{Bondi gauge metric}, we will assume the following falloffs in $u$ when $u\to \pm \infty$ \cite{Compere:2018ylh , Campiglia:2020qvc, Compere:2020lrt, Campiglia:2021bap,Donnay:2021wrk}: 
\begin{equation}
    N_{zz} = N_{zz}^{vac} + o(u^{-2}) ,\qquad C_{zz} = (u + C_\pm) N_{zz}^{vac} - 2 \partial^2 C_\pm + o(u^{-1})\, .
    \label{falloff in u}
\end{equation} Here, $C_{\pm}$ correspond to the values of the supertranslation field at $\mathscr{I}^+_\pm$ that encodes the displacement memory effect \cite{Strominger:2014pwa}. They transform as
\begin{equation}
    \delta_{(\mathcal{T}, \mathcal{Y}, \bar{\mathcal{Y}})} C_{\pm} =  \left(\mathcal{Y} \partial + \bar{\mathcal{Y}} \bar{\partial} - \frac{1}{2} \partial \mathcal{Y} - \frac{1}{2} \bar{\partial} \bar{\mathcal{Y}} \right) C_\pm + \mathcal{T}  \\.
    \label{transfo Cpm}
\end{equation} 
The vacuum News tensor $N_{zz}^{vac}$ \cite{Compere:2016jwb,Compere:2018ylh}, identified with the tracefree part of the Geroch tensor \cite{Geroch1977,Campiglia:2020qvc,Nguyen:2022zgs}, is given by
\begin{equation}
    N_{zz}^{vac} = \frac{1}{2}(\partial\varphi)^2 - \partial^2 \varphi \,,
    \label{Liouville stress tensor}
\end{equation}
in terms of the holomorphic superboost scalar field $\varphi (z)$. The latter encodes the refraction/velocity kick memory effects \cite{Compere:2018ylh}. We have
\begin{equation}
    \delta_{(\mathcal{T}, \mathcal{Y}, \bar{\mathcal{Y}})} \varphi = \mathcal{Y} \partial \varphi + \partial \mathcal{Y}\,,
\end{equation}
    so that
\begin{equation}
    \delta_{(\mathcal{T}, \mathcal{Y}, \bar{\mathcal{Y}})} N_{zz}^{vac} = (\mathcal{Y} \partial + 2 \partial \mathcal{Y}) N_{zz}^{vac} - \partial^3 \mathcal{Y}\,.
    \label{transfo Nvac}
\end{equation} One can check using \eqref{transfo CN}, \eqref{transfo Cpm} and \eqref{transfo Nvac} that the falloff conditions \eqref{falloff in u} are compatible with the action of extended BMS symmetries. Let us point out that these falloffs eliminate potentially interesting asymptotically flat solutions such as the gravitational tails producing terms $N_{zz} \sim \mathcal{O}(u^{-2})$ (see e.g. \cite{Blanchet:1987wq,Blanchet:1993ec,Blanchet:2020ngx}). These effects are potentially important if one wants to account for the $\mathcal{O}(\log \omega)$ corrections to the soft theorems arising at one-loop \cite{Sahoo:2018lxl,Laddha:2018vbn,Campiglia:2019wxe,Campiglia:2021bap}. However, in the present analysis the falloffs \eqref{falloff in u} are needed to have finite superrotation fluxes. Thus a description of $\mathcal{O}(\log \omega)$ corrections to the soft factors falls beyond the scope of the present analysis, mirroring and reproducing the results originally obtained by Bern, Davies and Nohle who assumed the absence of such terms in carrying out their work \cite{Bern:2014oka}. It would be interesting to explore if some renormalisation procedure could be applied in the present context to remove the divergences in $u$ and weaken the falloffs \eqref{falloff in u}. Notice that the latter imply that $N^{vac}_{zz}$ takes the same values at $\mathscr{I}^+_-$ and $\mathscr{I}^+_+$, which is consistent with vacuum Einstein equations. To allow some variations of $N^{vac}_{zz}$ and superboost transitions for a given solution, one would need to source the gravitational field with some exotic type of matter with falloffs $T_{AB} \sim \mathcal{O}(r)$.

The BMS surface charges obtained by using covariant phase space methods \cite{Lee:1990nz,Wald:1993nt,Iyer:1994ys,Barnich:2001jy} are non-integrable and non-conserved due to the presence of outgoing radiation \cite{Wald:1999wa,Barnich:2011mi}. Progress has been made recently in \cite{Compere:2018ylh,Compere:2020lrt,Ruzziconi:2020cjt,Donnay:2021wrk,Fiorucci:2021pha,Donnay:2022aba} to single out a preferred finite charge expression. Based on these works, we will use the BMS surface charges
\begin{equation}
    Q_\xi = \frac{1}{8\pi G} \int_{\mathcal{S}}  d^2z\,  [2\mathcal{T} \mathcal{M} + \mathcal{Y} \bar{\mathcal{N}} + \bar{\mathcal{Y}} \mathcal{N}]\,,
    \label{BMS surface charges}
\end{equation} where the supermomentum $\mathcal{M}$ and super angular momentum $\mathcal{N}$ are given by
\begin{equation}
\begin{split}
    \mathcal{M} &=   M + \frac{1}{8} ( C_{zz} N^{zz}+ C_{\bar{z}\bar{z}} N^{\bar{z}\bar{z}})\,,\\
    \mathcal{N} &=   N_{\bar{z}} - u  \bar{\partial}  \mathcal{M} + \frac{1}{4} C_{\bar{z}\bar{z}} \bar{\partial} C^{\bar{z}\bar{z}} + \frac{3}{16} \bar{\partial} (C_{zz}C^{zz}) \\
        &\quad+ \frac{u}{4} \bar{\partial} \Big[ \Big( \partial^2 - \frac{1}{2} N_{zz}   \Big)C^z_{\bar{z}} - \Big( \bar{\partial}^2 - \frac{1}{2} N_{\bar{z}\bar{z}}\Big)C^{\bar{z}}_z \Big]\,.
\end{split}    \label{BMS momenta}
\end{equation} The integration in \eqref{BMS surface charges} takes place on a cut $\mathcal{S}$ of $\mathscr{I}^+$ diffeomorphic to the one-punctured complex plane, so that the superrotations are globally well-defined. Integration on $\mathcal{S}$ can be defined using complexification of $(z, \bar{z})$ and residue integrals; see \cite{Barnich:2021dta,Donnay:2021wrk}. The momenta \eqref{BMS momenta} slightly differ from those considered in \cite{Donnay:2021wrk};\footnote{In \cite{Donnay:2021wrk}, the supermomentum aspect was taken to be $\mathcal{M}_{there} =   M + \frac{1}{8} ( C_{zz} N^{zz}_{vac}+ C_{\bar{z}\bar{z}} N^{\bar{z}\bar{z}}_{vac})$. Hence, this prescription differs from the one used here as $\mathcal{M}_{here} = \mathcal{M}_{there} + \frac{1}{8} ( C_{zz} \hat{N}^{zz}+ C_{\bar{z}\bar{z}} \hat{N}^{\bar{z}\bar{z}})$ where $\hat{N}_{zz} = N_{zz} - N_{zz}^{vac}$. Similarly, for the super angular momentum, we have $\mathcal{N}_{here} = \mathcal{N}_{there} - \frac{u}{4} \bar{\partial} ( \hat{N}_{zz} C^{zz})$.} however one can check that they lead to the same BMS fluxes discussed in \cite{Donnay:2021wrk} thanks to the falloffs \eqref{falloff in u}. The modifications are such that in the Newman-Penrose formalism \cite{Newman:1961qr, Newman:1962cia}, the BMS momenta \eqref{BMS momenta} simply read
\begin{equation}
   \mathcal{M} = -\frac{1}{2}(\Psi^0_2 + \bar{\Psi}^0_2) \,, \qquad \mathcal{N} = -\Psi^0_1 + u \bar{\partial}\Psi^0_2 \,.
    \label{momenta in NP}
\end{equation}
See appendix \ref{sec:Newman-Penrose formalism vs Metric formalism} for the conventions and the details of the translation to Newman-Penrose quantities.
These expressions coincide with the renormalised corner charge
aspects recently discussed in \cite{Freidel:2021qpz,Freidel:2021dfs,Freidel:2021ytz}. Let us emphasize that \eqref{momenta in NP} calls for the dual supermomentum $\tilde{\mathcal M} = - \frac{1}{2} (\Psi^0_2 - \bar{\Psi}^0_2)$ \cite{Godazgar:2018qpq,Kol:2019nkc,Oliveri:2020xls,Godazgar:2020gqd,Godazgar:2020kqd, Freidel:2021qpz}. Indeed, while $\mathcal{N}$ involves $u \bar{\partial} \Psi^0_2$, the supermomentum $\mathcal{M}$ implies only the real part of $\Psi^0_2$. It is suggestive to consider the following combination for the BMS charges,
\begin{equation}
\begin{split}
     Q_\xi^{full} &= -\frac{1}{8\pi G} \int_{\mathcal{S}}  d^2z\,  [2 \xi^u \Psi^0_2 + \mathcal{Y} \bar{\Psi}^0_1 + \bar{\mathcal{Y}} \Psi^0_1] \\
     &=\frac{1}{8\pi G} \int_{\mathcal{S}}  d^2z\,  [2\mathcal{T} (\mathcal{M} + i \tilde{\mathcal{M}}) + \mathcal{Y} \bar{\mathcal{N}} + \bar{\mathcal{Y}} \mathcal{N}]\,,
\end{split}
\end{equation} 
in agreement with the analysis of \cite{Godazgar:2018qpq,Kol:2019nkc,Oliveri:2020xls,Godazgar:2020gqd,Godazgar:2020kqd, Freidel:2021qpz}. In the following, we will not consider the contribution brought by $\tilde{\mathcal{M}}$ and restrict to the charge expression \eqref{BMS surface charges}. It would be interesting to investigate how $\tilde{\mathcal{M}}$ would impact the analysis presented in this paper. 

The integrated BMS fluxes are in turn given by
\begin{equation}
    F_{\xi} = \int_{-\infty}^{+\infty} du\, \partial_u Q_\xi = \int_{\mathcal{S}} d^2z \, [\mathcal{T} (\mathcal{P}_{hard}  + \mathcal{P}_{soft}) + \mathcal{Y} ( \bar{\mathcal{J}}_{hard} + \bar{\mathcal{J}}_{soft}) + \bar{\mathcal{Y}} ( \mathcal{J}_{soft} + \mathcal{J}_{hard} ) ],
    \label{BMS fluxes abstract}
\end{equation} where the explicit expressions for $\mathcal{P}_{hard/soft}$, $\mathcal{J}_{hard/soft}$ and $\bar{\mathcal{J}}_{hard/soft}$ can be found in \cite{Donnay:2021wrk}. There, a new split between soft and hard fluxes was proposed so that the latter form a representation of the BMS algebra separately. The expressions of these fluxes are rewritten in a more elegant form in equations \eqref{supertranslation fluxes} and \eqref{superrotation fluxes} in terms of natural variables of the radiative phase space to be introduced in Section \ref{sec:Radiative phase space with superrotations}. 

The centerless Virasoro algebra being part of the extended BMS algebra, it is also natural to introduce the notion of conformal fields. A primary conformal field $\phi_{h, \bar{h}}(z, \bar{z})$ of weights $(h, \bar{h})$ transforms under the action of superrotations as
\begin{equation}
    \delta_{(\mathcal{Y}, \bar{\mathcal{Y}})} \phi_{h, \bar{h}} = (\mathcal{Y} \partial + \bar{\mathcal{Y}} \bar{\partial} + h \partial \mathcal{Y} +\bar{h} \bar{\partial} \bar{\mathcal{Y}} ) \phi_{h, \bar{h}}\,.
    \label{def conformal field}
\end{equation} 
To make the covariance of the expressions under the superrotations manifest, it is useful to introduce the derivative operators~\cite{Barnich:2021dta,Campiglia:2020qvc,Donnay:2021wrk,Freidel:2021ytz}
\begin{equation}
    \begin{split}
       \mathscr{D}\phi_{h, \bar{h}} = [\partial - h \partial \varphi] \phi_{h, \bar{h}}\,,  \qquad 
       \bar{\mathscr{D}}\phi_{h, \bar{h}} =  [\bar{\partial} - \bar{h} \bar{\partial} \bar{\varphi}] \phi_{h, \bar{h}}\,,
    \end{split}  \label{derivative operators conformal}
\end{equation} which produce conformal fields of weights $(h+1, \bar{h})$ and $(h, \bar{h}+1)$, respectively. Notice that they also satisfy $[\mathscr{D}, \bar{\mathscr{D}}] \phi_{h, \bar{h}} = 0$.

\section{Extended radiative phase space}
\label{sec:Radiative phase space with superrotations}

In this section, we introduce hard and soft variables of the radiative phase space at $\mathscr{I}^+$ which are naturally suited when considering superrotations. We then construct the symplectic structure and show that the associated canonical generators reproduce the BMS fluxes \eqref{BMS fluxes abstract} of \cite{Donnay:2021wrk} that generate the variations on the phase space. In particular, we will observe that the hard and soft sectors of the phase space nicely factorise, which confirms the analysis of \cite{Campiglia:2021bap}.

We define $C^{(0)}_{zz}$, $\tilde{C}_{zz}$ and $\tilde{N}_{zz}$ through\footnote{Comparing with $\hat{N}_{zz}$ and $\hat{C}_{zz}$ defined in \cite{Donnay:2021wrk}, we have $\tilde{C}_{zz} + C^{(0)}_{zz} = \hat{C}_{zz}$ and $\tilde{N}_{zz} = \hat{N}_{zz}$.}
\begin{equation}
 \begin{split} 
&C_{zz}= u N^{vac}_{zz} + C^{(0)}_{zz} + \tilde C_{zz}\,, \qquad N_{zz} = N_{zz}^{vac} + \tilde{N}_{zz}\,, \\
&C^{(0)}_{zz}=-2 \mathscr{D}^2 C^{(0)}\, , \qquad C^{(0)}=\frac{1}{2}(C_++C_-) \,.
 \end{split} \label{defs fields}
\end{equation}
With these definitions, we have the following conditions at the corners of $\mathscr{I}^+$ as a consequence of \eqref{falloff in u},
\begin{equation}
\begin{split}
     &\tilde{C}_{zz}|_{\mathscr{I}^+_\pm} = \mp 2  \mathscr{D}^2 N^{(0)}\,, \qquad \tilde{N}_{zz}|_{\mathscr{I}^+_\pm} = 0\,, \qquad N^{(0)} = \frac{1}{2} (C_+ - C_- ) \,.
     \label{asymptotic values}
\end{split}
\end{equation} The leading soft News $\mathcal{N}^{(0)}_{zz}$ and subleading soft News $\mathcal{N}^{(1)}_{zz}$ are defined through
\begin{equation}
    \mathcal{N}^{(0)}_{zz} = \int_{-\infty}^{+\infty} du \, \tilde{N}_{zz} = -4 \mathscr{D}^2 N^{(0)}\,, \qquad \mathcal{N}^{(1)}_{zz} = \int_{-\infty}^{+\infty} du \,  u \tilde{N}_{zz} \,.
    \label{soft news}
\end{equation} The hard variables of the radiative phase space are given by
\begin{equation}
    \Gamma^{hard} = \{ \tilde{C}_{zz}, \tilde{N}_{zz}, \tilde{C}_{\bar{z}\bar{z}}, \tilde{N}_{\bar{z}\bar{z}} \} \,,
\end{equation} while the soft variables are 
\begin{equation}
\begin{split}
    \Gamma^{soft} &= \{  C^{(0)}, N^{(0)} , \mathcal{N}^{(1)}_{zz}, \mathcal{N}^{(1)}_{\bar{z}\bar{z}},  \varphi, \bar{\varphi}\,|\,\bar{\partial} \varphi = 0 = \partial \bar{\varphi}  \}\\
    &= \{ C^{(0)}_{zz}, C^{(0)}_{\bar{z}\bar{z}} , \mathcal{N}^{(0)}_{zz}, \mathcal{N}^{(0)}_{\bar{z}\bar{z}} , \Pi_{zz} , \Pi_{\bar{z}\bar{z}} , N_{zz}^{vac},  N_{\bar{z}\bar{z}}^{vac}\,|\\ 
    &\qquad\bar{\mathscr{D}}^2 C^{(0)}_{zz} = \mathscr{D}^2 C^{(0)}_{\bar{z}\bar{z}}, \bar{\mathscr{D}}^2 \mathcal{N}^{(0)}_{zz} = \mathscr{D}^2 \mathcal{N}^{(0)}_{\bar{z}\bar{z}}, \bar{\partial} N_{zz}^{vac} = 0 = \partial N_{\bar{z}\bar{z}}^{vac}  \}\,,
\end{split} \label{soft variables}
\end{equation} 
where the field $\Pi_{zz} = 2 \mathcal{N}^{(1)}_{zz} + C^{(0)} \mathcal{N}^{(0)}_{zz}$ corresponds to the symplectic partner of $N_{\bar{z}\bar{z}}^{vac}$ \cite{Campiglia:2021bap}. As we will explain below, the parametrisation in the second line of \eqref{soft variables} corresponds to Darboux coordinates for the symplectic form. Using \eqref{transfo CN}, \eqref{transfo Cpm} and \eqref{transfo Nvac}, one can deduce the transformations of the above phase space variables under extended BMS symmetries,
\begin{equation}
    \begin{split}
    &\delta_{(\mathcal{T}, \mathcal{Y}, \bar{\mathcal{Y}})} \tilde{C}_{zz} =  \left(\mathcal{Y} \partial + \bar{\mathcal{Y}} \bar{\partial} + \frac{3}{2} \partial \mathcal{Y} - \frac{1}{2} \bar{\partial} \bar{\mathcal{Y}} \right) \tilde{C}_{zz} + \left( \mathcal{T} + \frac{u}{2} (\partial \mathcal{Y} + \bar{\partial} \bar{\mathcal{Y}}) \right) \tilde{N}_{zz} \,, \\
    &\delta_{(\mathcal{T}, \mathcal{Y}, \bar{\mathcal{Y}})} \tilde{N}_{zz} =  \left(\mathcal{Y} \partial + \bar{\mathcal{Y}} \bar{\partial} + 2 \partial \mathcal{Y} \right) \tilde{N}_{zz} + \left(\mathcal{T} + \frac{u}{2} (\partial \mathcal{Y} + \bar{\partial} \bar{\mathcal{Y}}) \right) \partial_u \tilde{N}_{zz} \,, \\
       &\delta_{(\mathcal{T}, \mathcal{Y}, \bar{\mathcal{Y}})} {C}^{(0)}_{zz} =  \left(\mathcal{Y} \partial + \bar{\mathcal{Y}} \bar{\partial} + \frac{3}{2} \partial \mathcal{Y} - \frac{1}{2} \bar{\partial} \bar{\mathcal{Y}} \right) {C}^{(0)}_{zz} -2 \mathscr{D}^2 \mathcal{T} \,, \\
       &\delta_{(\mathcal{T}, \mathcal{Y}, \bar{\mathcal{Y}})} \mathcal{N}^{(0)}_{zz} = \left(\mathcal{Y} \partial + \bar{\mathcal{Y}} \bar{\partial} + \frac{3}{2} \partial \mathcal{Y} - \frac{1}{2} \bar{\partial} \bar{\mathcal{Y}} \right) \mathcal{N}^{(0)}_{zz} \,,\\
       &\delta_{(\mathcal{T}, \mathcal{Y}, \bar{\mathcal{Y}})} \Pi_{zz} =   \left(\mathcal{Y} \partial + \bar{\mathcal{Y}} \bar{\partial} + \partial \mathcal{Y} - \bar{\partial} \bar{\mathcal{Y}} \right) \Pi_{zz} - \mathcal{T} \mathcal{N}^{(0)}_{zz} \,, \\
       &\delta_{(\mathcal{T}, \mathcal{Y}, \bar{\mathcal{Y}})} N_{zz}^{vac} = (\mathcal{Y} \partial + 2 \partial \mathcal{Y}) N_{zz}^{vac} - \partial^3 \mathcal{Y} \,,
    \end{split} 
    \label{variations BMS}
\end{equation}
together with the complex conjugate relations. In particular, the soft variables ${C}^{(0)}_{zz}$, $\mathcal{N}^{(0)}_{zz}$ and $\Pi_{zz}$ are conformal fields in the sense of \eqref{def conformal field}. 

A symplectic form $\Omega$ on the extended radiative phase space at $\mathscr{I}^+$ was proposed recently by Campiglia and Laddha \cite{Campiglia:2021bap}. It factorises into hard and soft sectors,
\begin{equation}
\begin{split}
    &\Omega =  \Omega^{hard} + \Omega^{soft} \,,\\
    &\Omega^{hard} = \frac{1}{32\pi G} \int_{\mathscr{I}^+} du d^2 z \left[ \delta \tilde{N}_{zz} \wedge \delta \tilde{C}_{\bar{z}\bar{z}}  + c.c.\right] \,, \\
    &\Omega^{soft} = \frac{1}{32\pi G} \int_{\mathcal{S}}  d^2 z \left[\delta \mathcal{N}^{(0)}_{zz} \wedge \delta C^{(0)}_{\bar{z}\bar{z}}  + \delta \Pi_{zz}  \wedge \delta N_{\bar{z}\bar{z}}^{vac} +c.c.   \right] \,.
\end{split}
\label{symplectic structure full}
\end{equation} Using the definitions in \eqref{defs fields} and \eqref{soft news} and integration by parts, it can be related to the Ashtekar-Streubel (AS) symplectic structure \cite{Ashtekar:1978zz,Ashtekar:1981bq,Ashtekar:1981sf}
\begin{equation}
    \Omega_{AS} = \frac{1}{32\pi G} \int_{\mathscr{I}^+} du d^2z \left[\delta N_{zz} \wedge \delta C_{\bar{z}\bar{z}} + c.c. \right]\,,
\end{equation} 
through
\begin{equation}
    \Omega = \Omega_{AS} + \frac{1}{32\pi G} \int_{\mathcal{S}}  d^2 z  \left[ \delta (C^{(0)} \mathcal{N}^{(0)}_{zz}) \wedge \delta N_{\bar{z}\bar{z}}^{vac} + c.c. \right] \,.
\end{equation} 
We note that the difference between $\Omega$ and $\Omega_{AS}$ can be interpreted as a corner ambiguity of the symplectic form \cite{Lee:1990nz,Wald:1993nt,Iyer:1994ys} (see also \cite{Compere:2018aar,Ruzziconi:2019pzd,Fiorucci:2021pha} for reviews):
\begin{equation}
\begin{split}
    &\Omega = \Omega_{AS} + \int_{\mathscr{I}^+} d \delta Y, \\
    &\delta Y =   \frac{1}{32\pi G}\left[  \delta (C^{(0)} \tilde{C}_{zz}) \wedge \delta N_{\bar{z}\bar{z}}^{vac} + c.c. \right] dz \wedge d\bar{z}\,.
\end{split}
\end{equation} Here, $Y$ is a one-form on the phase space and a two-form on the spacetime. This ambiguity can be derived from the variational principle \cite{Papadimitriou:2005ii,Compere:2008us,Harlow:2019yfa,Freidel:2020xyx,Fiorucci:2020xto,Freidel:2021fxf,Chandrasekaran:2021vyu} by adding to the action a boundary term of the form 
\begin{equation}
S_b =- \frac{1}{32\pi G} \int_{\mathscr{I}^+} du d^2z \, [ C^{(0)} \tilde{N}_{zz} N_{\bar{z}\bar{z}}^{vac} + c.c. ]\,.
\end{equation} 

From the symplectic structure \eqref{symplectic structure full}, one can read the non-vanishing Poisson brackets\footnote{The sign convention to define the Poisson bracket associated to the symplectic structure \eqref{symplectic structure full} is chosen to match with the bracket introduced in \cite{Donnay:2021wrk}. It is such that for two functions $f$, $g$ on the phase space, we have $\{ f, g \} = i_{X_f} i_{X_g} \Omega$, where the Hamiltonian vector field $X_f$ satisfies $i_{X_f} \Omega = \delta f$.}
\begin{equation}
    \begin{split}  
        &\{ \tilde C_{zz}(u), \tilde{C}_{\bar{w}\bar{w}}(u')  \} = -8\pi G\, \delta^{(2)} (z-w) \text{sgn} (u-u') \,,\\
    & \{ \mathcal{N}^{(0)}_{zz} , C^{(0)}_{\bar{w}\bar{w}}\} = -16\pi G\,  \delta^{(2)}(z-w) \,,\\
     & \{ \Pi_{zz}, N_{\bar{w}\bar{w}}^{vac}  \} =- 16 \pi G\, \delta^{(2)}(z-w)\,.
    \end{split}
    \label{non vanishing brackets}
\end{equation} In particular, using $\partial_u \text{sgn}(u) = 2 \delta (u)$, the first bracket implies
\begin{equation}
    \begin{split}
        &\{ \tilde{N}_{zz}(u), \tilde{C}_{\bar{w}\bar{w}}(u') \} = -16\pi G\,  \delta^{(2)} (z-w) \delta (u-u')\,, \\
        &\{ \tilde{N}_{zz}(u), \tilde{N}_{\bar{w}\bar{w}}(u') \} = 16\pi G\,  \delta^{(2)} (z-w) \partial_u \delta (u-u')\,.
    \end{split}
\end{equation} 

The BMS fluxes are obtained by contracting the symplectic structure \eqref{symplectic structure full} with the Hamiltonian vector fields generating BMS symmetries. Using \eqref{variations BMS}, we have
\begin{equation}
    i_{\delta_{(\mathcal{T}, \mathcal{Y}, \bar{\mathcal{Y}})}} \Omega^{soft} = \delta F^{soft}_{(\mathcal{T}, \mathcal{Y}, \bar{\mathcal{Y}})}\,, \qquad i_{\delta_{(\mathcal{T}, \mathcal{Y}, \bar{\mathcal{Y}})}} \Omega^{hard} = \delta F^{hard}_{(\mathcal{T}, \mathcal{Y}, \bar{\mathcal{Y}})}\,,
    \label{canonical generators}
\end{equation} where the supertranslation fluxes are given by
\begin{equation}
\begin{split}
     &F_{\mathcal{T}}^{hard} = - \frac{1}{16\pi G} \int_{\mathscr{I}^+} du d^2z\, \mathcal{T} \left[ \tilde{N}_{zz}  \tilde{N}_{\bar{z}\bar{z}} \right] \,,\\
     &F_{\mathcal{T}}^{soft} = \frac{1}{8\pi G} \int_{\mathcal{S}} d^2 z\, \mathcal{T} \left[ \mathscr{D}^2\mathcal{N}^{(0)}_{\bar{z}\bar{z}} \right]\,,
\end{split} \label{supertranslation fluxes}
\end{equation}
and the superrotation fluxes by
\begin{equation}
\begin{split}
     &F_{\mathcal{Y}}^{hard} = \frac{1}{16\pi G} \int_{\mathscr{I}^+} du d^2 z\,  {\mathcal{Y}} \left[\frac{3}{2}\tilde C_{zz} \partial \tilde N_{\bar z \bar z}+ \frac{1}{2}\tilde N_{\bar z \bar z} \partial \tilde C_{zz}+\frac{u}{2} \partial (\tilde N_{zz} \tilde N_{\bar z \bar z})\right] \,,\\
     &F_{\mathcal{Y}}^{soft} = \frac{1}{16\pi G}\int_\mathcal{S} d^2 z\,  \mathcal{Y}  \left[-{\mathscr D}^3 \mathcal N_{\bar z \bar z}^{(1)} + \frac{3}{2} C^{(0)}_{zz} \mathscr{D} \mathcal{N}^{(0)}_{\bar z \bar z}+ \frac{1}{2}\mathcal{N}^{(0)}_{\bar z \bar z} \mathscr{D} C_{zz}^{(0)}  \right]\,,
\end{split}
\label{superrotation fluxes}
\end{equation} together with the complex conjugate expressions associated with $\bar{\mathcal{Y}}$. Using the definitions \eqref{defs fields} and \eqref{soft news}, it is straightforward to show that the BMS fluxes \eqref{supertranslation fluxes} and \eqref{superrotation fluxes} precisely reproduce the hard and soft fluxes provided in \cite{Donnay:2021wrk} and have the right behavior under extended BMS transformations. More precisely, the hard and soft BMS momentum fluxes appearing in \eqref{BMS fluxes abstract} transform separately in the coadjoint representation \cite{Barnich:2021dta,Donnay:2021wrk},
\begin{equation}
\badat{2}
\label{delta P and J}
        &\delta_{(\mathcal{T}, \mathcal{Y}, \bar{\mathcal{Y}})} \mathcal{P}_{hard/soft} = \Big[\mathcal{Y} \partial + \bar{\mathcal{Y}} \bar{\partial} + \frac{3}{2} \partial \mathcal{Y} + \frac{3}{2} \bar{\partial} \bar{\mathcal{Y}} \Big]  \mathcal{P}_{hard/soft}\,, \\
        &\delta_{(\mathcal{T}, \mathcal{Y}, \bar{\mathcal{Y}})} \mathcal{J}_{hard/soft} = \Big[ \mathcal{Y} \partial + \bar{\mathcal{Y}} \bar{\partial} + 2 \bar{\partial} \bar{\mathcal{Y}} + \partial  \mathcal{Y} \Big] \mathcal{J}_{hard/soft} \\
        &\qquad\qquad\qquad\qquad+ \frac{1}{2} \mathcal{T} \bar{\partial} \mathcal{P}_{hard/soft} + \frac{3}{2} \bar{\partial} \mathcal{T} \mathcal{P}_{hard/soft} \,.
\eadat
\end{equation} In particular, the last two terms appearing in the soft flux \eqref{superrotation fluxes},
\begin{equation}
   \frac{1}{16\pi G}\int_\mathcal{S} d^2 z\,  \mathcal{Y} \left[ \frac{3}{2} C^{(0)}_{zz} \mathscr{D} \mathcal{N}^{(0)}_{\bar z \bar z}+ \frac{1}{2}\mathcal{N}^{(0)}_{\bar z \bar z} \mathscr{D} C_{zz}^{(0)} \right]\,, 
   \label{one loop terms}
\end{equation} were introduced in \cite{Donnay:2021wrk} in order to reproduce the correct inhomogeneous term in the transformation of the super angular momentum flux under supertranslation, namely the terms linear in $\mT$ in \eqref{delta P and J}.\footnote{In Newtonian mechanics, the angular momentum $\Vec{L}$ transforms inhomogeneously under a translation $\Vec{\alpha}$ in terms of the cross product between the translation and the linear momentum $\Vec{p}$:  $\delta_{{\alpha}} \Vec{L} = \Vec{\alpha} \times \Vec{p}$. In \cite{Donnay:2021wrk} terms involving $C_-$ were introduced in the split between hard and soft super angular momentum fluxes to reproduce the generalization of this transformation to supertranslations. It was also noted in footnote~8 of that paper that the same terms with $C_+$ instead of $C_-$ could have been chosen since both fields transform in the same way under extended BMS symmetries. Here we use their weighted sum $C^{(0)}$ instead.} As we will argue in Section \ref{sec:Subleading soft theorem}, the terms \eqref{one loop terms} are crucial to recover the one-loop corrected subleading soft theorem and corresponding conformal Ward identities. 

For practical purposes, it turns out to be useful to provide an alternative expression for the superrotation fluxes \eqref{superrotation fluxes}. First, integrating by parts with respect to $u$ and using the asymptotic values of the fields \eqref{asymptotic values} to cancel boundary terms, the hard flux can be rewritten as 
\begin{equation}
    F_{\mathcal{Y}}^{hard} = \frac{1}{16\pi G} \int_{\mathscr{I}^+} du d^2 z\,  {\mathcal{Y}} \left[-\frac{3}{2}\tilde N_{zz} \partial \tilde C_{\bar z \bar z}- \frac{1}{2}\tilde C_{\bar{z}\bar{z}} \partial \tilde N_{z z} +\frac{u}{2} \partial (\tilde N_{zz} \tilde N_{\bar z \bar z})\right]\,.
\end{equation} Furthermore, integrating by parts with respect to $z$ and using the electricity conditions \eqref{asymptotic values} and \eqref{soft news} for $\mathcal{N}^{(0)}_{\bar{z}\bar{z}}$ and $C^{(0)}_{zz}$, the soft flux can be rewritten as
\begin{align}
F_{\mathcal{Y}}^{soft} &= \frac{1}{16\pi G}\int_{\mathcal{S}} d^2 z\,  \mathcal{Y}  \left[-{\mathscr D}^3 (\mathcal N_{\bar z \bar z}^{(1)} + C^{(0)} \mathcal{N}^{(0)}_{\bar z \bar z} ) - \frac{3}{2} \mathcal{N}^{(0)}_{zz} \mathscr{D} C^{(0)}_{\bar z \bar z}- \frac{1}{2} C^{(0)}_{\bar z \bar z} \mathscr{D} \mathcal{N}_{zz}^{(0)}  \right] \label{FY soft}\\
&= \frac{1}{16\pi G}\int_{\mathcal{S}} d^2 z\,  \mathcal{Y}  \left[-{\mathscr D}^3 (\mathcal N_{\bar z \bar z}^{(1)} + C^{(0)} \mathcal{N}^{(0)}_{\bar z \bar z} ) +3 \bar{\mathscr{D}}^2\mathcal{N}^{(0)}_{zz} \mathscr{D} C^{(0)}+ C^{(0)} \mathscr{D}\bar{\mathscr{D}}^2 \mathcal{N}_{zz}^{(0)}  \right] \,. \nonumber
\end{align}

It can be shown that the BMS fluxes generate the transformations on the radiative phase space, namely
\begin{equation}
    \begin{split}
        &\{ F^{hard}_{(\mathcal{T}, \mathcal{Y}, \bar{\mathcal{Y}})} , \tilde{C}_{zz}  \} =  \delta_{(\mathcal{T}, \mathcal{Y}, \bar{\mathcal{Y}})} \tilde{C}_{zz} , \quad \{ F^{hard}_{(\mathcal{T}, \mathcal{Y}, \bar{\mathcal{Y}})} , \tilde{N}_{zz}  \} =  \delta_{(\mathcal{T}, \mathcal{Y}, \bar{\mathcal{Y}})} \tilde{N}_{zz} , \\
        &\{ F^{soft}_{(\mathcal{T}, \mathcal{Y}, \bar{\mathcal{Y}})} , C^{(0)}_{zz} \}  =   \delta_{(\mathcal{T}, \mathcal{Y}, \bar{\mathcal{Y}})} C^{(0)}_{zz} , \quad \{ F^{soft}_{(\mathcal{T}, \mathcal{Y}, \bar{\mathcal{Y}})} , N^{(0)}_{zz} \}  =   \delta_{(\mathcal{T}, \mathcal{Y}, \bar{\mathcal{Y}})} N^{(0)}_{zz}, \\
        &\{ F^{soft}_{(\mathcal{T}, \mathcal{Y}, \bar{\mathcal{Y}})} , \Pi_{zz} \}  =   \delta_{(\mathcal{T}, \mathcal{Y}, \bar{\mathcal{Y}})} \Pi_{zz}, \\
         &\{ F^{soft}_{(\mathcal{T}, \mathcal{Y}, \bar{\mathcal{Y}})} , \tilde{C}_{zz}  \} = 0 , \quad \{ F^{soft}_{(\mathcal{T}, \mathcal{Y}, \bar{\mathcal{Y}})} , \tilde{N}_{zz}  \} = 0 \,, \\
         &\{ F^{hard}_{(\mathcal{T}, \mathcal{Y}, \bar{\mathcal{Y}})} , C^{(0)} \}  =  0\,, \quad \{ F^{hard}_{(\mathcal{T}, \mathcal{Y}, \bar{\mathcal{Y}})} , N^{(0)} \}  =  0\,, \quad \{ F^{hard}_{(\mathcal{T}, \mathcal{Y}, \bar{\mathcal{Y}})} , \Pi_{zz} \}  = 0\,.
    \end{split}
\end{equation} As a corollary, we have
\begin{equation}
    \{ F_{(\mathcal{T}_1, \mathcal{Y}_1, \bar{\mathcal{Y}}_1)}, F_{(\mathcal{T}_2, \mathcal{Y}_2, \bar{\mathcal{Y}}_2)} \} = \delta_{(\mathcal{T}_1, \mathcal{Y}_1, \bar{\mathcal{Y}}_1)} F_{(\mathcal{T}_2, \mathcal{Y}_2, \bar{\mathcal{Y}}_2)} \,,
\end{equation} which reproduces the bracket postulated in \cite{Donnay:2021wrk}. Using this bracket, it was shown there that the BMS fluxes represent the BMS algebra without central extension,
\begin{equation}
    \{ F_{(\mathcal{T}_1, \mathcal{Y}_1, \bar{\mathcal{Y}}_1)}, F_{(\mathcal{T}_2, \mathcal{Y}_2, \bar{\mathcal{Y}}_2)} \} =  - F_{[(\mathcal{T}_1, \mathcal{Y}_1, \bar{\mathcal{Y}}_1),(\mathcal{T}_2, \mathcal{Y}_2, \bar{\mathcal{Y}}_2)]}\, .
\end{equation} Furthermore, the factorisation of the symplectic form between hard and soft sectors implies
\begin{equation}
    \{ F^{soft}_{(\mathcal{T}_1, \mathcal{Y}_1, \bar{\mathcal{Y}}_1)} , F^{hard}_{(\mathcal{T}_2, \mathcal{Y}_2, \bar{\mathcal{Y}}_2)}\} = 0 \,, \qquad \{ F^{soft/hard} _{(\mathcal{T}_1 , \mathcal{Y}_1, \bar{\mathcal{Y}}_1)} , F_{(\mathcal{T}_2, \mathcal{Y}_2, \bar{\mathcal{Y}}_2)}^{soft/hard}   \}  = - F_{[(\mathcal{T}_1, \mathcal{Y}_1, \bar{\mathcal{Y}}_1), (\mathcal{T}_2, \mathcal{Y}_2, \bar{\mathcal{Y}}_2)]}^{soft/hard} \,,
    \label{factorization hard soft}
\end{equation} 
which confirms the suggestion made in \cite{Donnay:2021wrk}. Thus, the hard and soft fluxes separately give a representation of the extended BMS algebra. 

\section{Leading soft theorem and soft factorisation}
\label{sec:Leading soft theorem and soft factorisation}

In this section we review the leading soft graviton theorem and the soft factorisation of the $\mathcal{S}$-matrix. Some of their implications will be of direct use in the next section when discussing the one-loop corrections to the subleading soft graviton theorem.

We will treat both cases where external hard particles can be massless or massive. As usual, we adopt a parametrisation of massless momenta and polarisation vectors in terms of energy and stereographic coordinates on the celestial sphere (see e.g. \cite{Himwich:2020rro})
\begin{equation}
\begin{split}
q^\mu(\omega,z,\bz)&=\omega \left(1+z \bz\,,z+\bz\,,-i(z-\bz)\,, 1-z \bz \right)\equiv \omega\, \hat q^\mu(z,\bz)\,,\\
\varepsilon^+_\mu(z,\bz)&=\frac{1}{\sqrt{2}} \left(- \bz,1,-i,-\bz\right)\,,\\
\varepsilon^-_\mu(z,\bz)&=\frac{1}{\sqrt{2}} \left(-z,1,i,-z\right)\,.
\end{split}
\end{equation}
The two polarisation tensors of the graviton are constructed as $\varepsilon^\pm_{\mu\nu}=\varepsilon^\pm_\mu\, \varepsilon^\pm_\nu$ and they satisfy the relations
\begin{equation}
\varepsilon_\mu^+(z,\bz)=\frac{1}{\sqrt{2}}\, \partial_z \hat q_\mu(z,\bz)\,, \qquad \varepsilon_\mu^-(z,\bz)=\frac{1}{\sqrt{2}}\, \partial_{\bz} \hat q_\mu(z,\bz)\,.
\end{equation}
The inner product of two massless momenta is then simply given by
\begin{equation}
\hat q_1 \cdot \hat q_2=-2\, |z_{12}|^2\,.
\end{equation}
The parametrisation of massive momenta is slightly more involved as it refers to coordinates $(\rho,z,\bz)$ on the three-dimensional hyperboloid describing the directions of approach to timelike infinity $i^+$. We give it explicitly in appendix~\ref{app:massive}.

The leading soft graviton theorem is known since the work of Weinberg \cite{Weinberg:1965nx}. Consider the scattering amplitude $\mathcal{M}_n(p_1,...,p_n)$ of $n$ hard massless particles with momenta $p_i$'s. The leading soft graviton theorem states that the amplitude $\mathcal{M}_{n+1}(q,p_1,...,p_n)$ containing an additional external graviton with momentum $q=\omega \hat{q}$ and polarisation $\varepsilon^\pm_{\mu\nu}(\hat q)$, satisfies 
\begin{equation}
\label{leading soft theorem}
\lim_{\omega \to 0} \omega\, \mathcal{M}_{n+1}(\omega \hat q,p_1,...,p_n)=\hat S_n^{(0)\pm}\mathcal{M}_n(p_1,...,p_n)\,, 
\end{equation}
where the (normalised) soft factor is given by
\begin{equation}
\label{S0}
\hat S^{(0) \pm}_n=\frac{\kappa}{2} \sum_{i=1}^n \frac{p_i^\mu p_i^\nu \varepsilon^\pm_{\mu\nu}(\hat q)}{p_i \cdot \hat{q}}\,, \qquad \kappa^2=32\pi G\,.
\end{equation}
For ease of notation, we have taken all particles to be outgoing without loss of generality. In anticipation of the discussion that follows, we already point out that Weinberg's soft theorem can be written in terms of the insertion of the soft graviton operator $\mNzero_{zz}$ into the $\mathcal{S}$-matrix elements as
\begin{equation}
\label{N0 insertion}
\outstate \mNzero_{zz} \mathcal{S} \instate=-\frac{\kappa}{16\pi}\, \hat S^{(0)+}_n \outstate \mathcal{S} \instate\,.
\end{equation}
Indeed, $\mNzero_{zz}$ can be decomposed in terms of creation and annihilation operators of the graviton field \cite{He:2014laa},
\begin{equation}
\label{N0 creation}
\mNzero_{zz}=-\frac{\kappa}{16\pi}  \lim_{\omega \to 0} \omega \left[a_+^{\text{out}}(\omega \hat{q})+a_-^{\text{out}}(\omega \hat{q})^\dagger  \right]\,, 
\end{equation}
such that \eqref{N0 insertion} is equivalent to
\begin{equation}
\lim_{\omega \to 0} \omega\, \outstate a^{\text{out}}_+(\omega \hat q) \mathcal{S} \instate = \hat S^{(0)+}_n \outstate \mathcal{S} \instate\,.
\end{equation}
This is indeed Weinberg's soft graviton theorem \eqref{leading soft theorem}.

More recently Strominger, He, Lysov and Mitra showed that the leading soft theorem is equivalent to the conservation of BMS charges across spatial infinity, i.e., from $\scri^-_+$ to $\scri^+_-$ \cite{Strominger:2013jfa,He:2014laa},
\begin{equation}
\label{charge conservation}
Q(\scri^+_-)=Q(\scri^-_+)\,.
\end{equation}
To justify this conservation law, Strominger argued that antipodal matching conditions must be imposed on the field and symmetry parameters for the gravitational scattering problem to be well-defined \cite{Strominger:2013jfa}. It turns out that such antipodal matching relations and the resulting charge conservation follow directly from regularity of null infinity  \cite{Herberthson:1992gcz,Troessaert:2017jcm,Henneaux:2018cst,Henneaux:2018hdj,Prabhu:2019fsp,Prabhu:2021cgk,Mohamed:2021rfg,Capone:2022gme}. In spacetimes where charges vanish in the limit to $\scri^+_+$ and $\scri^-_-$, the conserved charges are also equal to the BMS fluxes passing through $\scri$,
\begin{equation}
Q(\scri^+_-)=-F(\scri^+)\,, \qquad Q(\scri^-_+)=F(\scri^-)\,.
\end{equation}
Therefore, invariance of the  $\mathcal{S}$-matrix under BMS symmetries reads \cite{Strominger:2013jfa}
\begin{equation}
    \left[ Q, \mathcal{S} \right] = 0\,,
\end{equation}
or equivalently
\begin{equation}
\label{FS-SF}
\outstate F(\scri^+) \mathcal{S}+ \mathcal{S} F(\scri^-) \instate =0\,.
\end{equation}
To recover the soft theorem, one considers the soft and hard pieces of the fluxes separately. The hard part of the flux is used to generate the BMS transformation of the hard external wavefunctions,
\begin{equation}
F^{hard}_{\mT} |\text{out}\rangle = -i \delta_{\mT} |\text{out}\rangle\,,
\end{equation}
while the soft part of the flux inserts the soft graviton field $\mNzero_{zz}$ as given in \eqref{N0 creation}. Putting this together one recovers the leading soft theorem in the form \eqref{N0 insertion}. Note that the treatment for massive external particles was given in \cite{Campiglia:2015kxa}.

Another important and related feature of the $\mathcal{S}$-matrix is its factorisation into soft and finite pieces \cite{Weinberg:1965nx,Naculich:2011ry,White:2011yy},
\begin{equation}
\label{soft facorisation}
\mathcal{M}=\mathcal{M}_{\text{soft}}\, \mathcal{M}_{\text{finite}}\,,
\end{equation}
where the soft factor $\mathcal{M}_{\text{soft}}$ carries all dependencies on the infrared regulator $\epsilon$,\footnote{In dimensional regularisation $d=4-2\epsilon$.} and takes the simple form
\begin{equation}
\mathcal{M}_{\text{soft}}=\exp \left[\frac{\hbar}{\epsilon} \frac{\kappa^2}{(8\pi)^2} \sum_{i,j=1}^n p_i \cdot p_j \ln \frac{p_i \cdot p_j}{\mu}\right]\,.
\end{equation}
Here $\mu$ is an immaterial parameter needed for dimensional reasons, and momentum conservation can be used to show that the above expression does not actually depend on its value. The dependence on $\epsilon$ is problematic however, as it makes the amplitude \eqref{soft facorisation} vanish in the limit $\epsilon \to 0$. In order to obtain meaningful predictions, one usually considers inclusive cross-sections with an arbitrary number of external soft gravitons. The Weinberg poles exponentiate precisely in such a way that they cancel the infrared soft factor $\mathcal{M}_{\text{soft}}$. 

This factorisation has motivated a factorisation of the external wavefunctions themselves \cite{Himwich:2020rro,Arkani-Hamed:2020gyp}. In the language of celestial holography, amplitudes are viewed as a correlators of a two-dimensional CFT living on the celestial sphere,
\begin{equation}
\mathcal{M}_n=\langle \text{out} | \mathcal{S} |\text{in} \rangle\equiv \langle \mO_1\, ...\, \mO_n \rangle\,,
\end{equation}
where the insertion of an operator $\mathcal{O}_i$ represents the hard external particle with momentum $p_i$. Factorisation of the external wavefunctions translates into factorisation of these operators. For massless particles it is assumed to take the form
\begin{equation}
\mO_\omega(z,\bz)=\mW_\omega(z,\bz)\, \tilde \mO_\omega(z,\bz)\,, \qquad \mW_\omega(z,\bz)=e^{i\omega C^{(0)}(z,\bz)}\,,
\end{equation}
where $\tilde \mO_\omega$ is invariant under supertranslations. The case of massive particles is treated in appendix~\ref{app:massive}. Further assuming that the two types of operators do not interact, the correlation functions factorise nicely,
\begin{equation}
\langle \mO_1\, ...\, \mO_n \rangle=\langle \mW_1\, ...\, \mW_n \rangle\, \langle \tilde \mO_1\, ...\, \tilde \mO_n \rangle\,,
\end{equation}
such that one makes the identifications
\begin{equation}
\mathcal{M}_{\text{soft}}=\langle \mW_1\, ...\, \mW_n \rangle\,, \qquad  \mathcal{M}_{\text{finite}}=\langle \tilde \mO_1\, ...\, \tilde \mO_n \rangle\,.
\end{equation}
Note that $\mathcal{M}_{\text{soft}}$ can also be written in terms of the celestial two-point correlator of the supertranslation Goldstone mode \cite{Himwich:2020rro,Arkani-Hamed:2020gyp,Nguyen:2021ydb},
\begin{equation}
\mathcal{M}_{\text{soft}}=\langle \mW_1\, ...\, \mW_n \rangle = \exp \left[-\frac{1}{2} \sum_{i \neq j}^n \eta_i \eta_j \omega_i \omega_j \langle C^{(0)}(z_i,\bz_i) C^{(0)}(z_j,\bz_j) \rangle\right]\,, 
\end{equation}
with
\begin{equation}
\label{CC correlator}
\langle C^{(0)}(z_i,\bz_i) C^{(0)}(z_j,\bz_j) \rangle =\frac{\hbar}{\epsilon}\frac{\kappa^2}{(4\pi)^2}\, |z_{ij}|^2 \ln |z_{ij}|^2 \,.
\end{equation}
In this description the supertranslation Goldstone mode is fully responsible for the infrared divergences of the $\mathcal{S}$-matrix. 

We can use the above structure to derive the effect of an insertion of the supertranslation mode $C^{(0)}$ into the $\mathcal{S}$-matrix. Indeed, we easily find
\begin{align}
\langle C^{(0)}(z, \bar{z}) \mW_1\, ...\, \mW_n \rangle&=-i \partial_\omega \langle \mW_{\omega}(z, \bar{z})\, \mW_1\, ...\, \mW_n \rangle \big|_{\omega=0}=- \frac{i\kappa^2}{\epsilon} \hat \sigma_{n+1}'\langle \mW_1\, ...\, \mW_n \rangle\,,
\end{align}
with
\begin{equation}
\hat \sigma_{n+1}'\equiv \frac{\hbar}{2(4\pi)^2} \sum_{i=1}^n (p_i \cdot \hat q) \ln \frac{p_i \cdot \hat q}{\mu}\,.
\end{equation}
Because the fields $\tilde \mO_i$ and $\mW_j$ do not interact, this also implies the $\mathcal{S}$-matrix identity
\begin{equation}
\label{C0 insertion}
\langle \text{out} | C^{(0)}(z, \bar{z}) \mathcal{S} |\text{in} \rangle=-\frac{i\kappa^2}{\epsilon}\, \hat \sigma_{n+1}' \langle \text{out} | \mathcal{S} |\text{in} \rangle \,.
\end{equation}
This identity is equally valid when the momenta $p_i$'s are massless or massive. The demonstration for the case of massive particles is given in appendix~\ref{app:massive}.

\section{Subleading soft theorem}
\label{sec:Subleading soft theorem}

We now come to the object of main interest in this work, namely the subleading soft theorem including loop contributions. The first evidence for the existence of a universal soft factor $S^{(1)}_n$ at subleading order in the energy $\omega$ of the soft graviton was provided by Cachazo and Strominger \cite{Cachazo:2014fwa}. This tree-level subleading soft theorem can be written
\begin{equation}
\label{tree level subleading theorem}
\lim_{\omega \to 0} \left(1+\omega \partial_\omega \right) \mathcal{M}^{\text{tree}}_{n+1}(\omega \hat q,p_1,...,p_n)=S^{(1)\pm}_n\, \mathcal{M}^{\text{tree}}_n(p_1,...,p_n)\,,
\end{equation}
where the role of $(1+\omega \partial_\omega)$ is to project out Weinberg's pole associated with the leading soft factor, and where the subleading soft factor is given in terms of the total (spin + orbital) angular momentum operator,
\begin{equation}
S^{(1)\pm}_n=-\frac{i \kappa}{2} \sum_{i=1}^n \frac{p_i^\mu\,  \varepsilon_{\mu\nu}^\pm(\hat q)\, q_\lambda}{p_i \cdot q}\, J_i^{\lambda\nu}\,.
\end{equation}
Following this new discovery, an equivalence between the tree-level subleading soft theorem and conservation of superrotation charges was also provided \cite{Kapec:2014opa}. The demonstration of this equivalence is completely analogous to the one between leading soft theorem and conservation of supertranslation charges, which we briefly reviewed in the last section. As a first step, one notes that the subleading soft theorem \eqref{tree level subleading theorem} can be written as an identity for the insertion of $\mNone_{\bz \bz}$,
\begin{equation}
\label{N1 insertion}
\outstate \mathcal{N}^{(1)}_{\bz \bz} \mathcal{S} \instate = \frac{i\kappa}{16\pi}\, S^{(1)-}_n \outstate \mathcal{S} \instate\,.
\end{equation}
Indeed the expression of $\mNone_{\bz \bz}$ in terms of creation and annihilation operators is given by
\begin{equation}
\mathcal{N}^{(1)}_{\bz \bz}=\frac{i \kappa}{16\pi} \lim_{\omega \to 0} \left(1+\omega \partial_\omega \right)\left[a^{\out}_-(\omega \hat q)  - a^{\out}_+(\omega \hat q)^\dagger \right]\,,
\end{equation}
such that \eqref{N1 insertion} becomes
\begin{equation}
\lim_{\omega \to 0} \left(1+\omega \partial_\omega\right) \outstate a^{\out}_-(\omega \hat q) \mathcal{S} \instate = S^{(1)-}_n \outstate \mathcal{S} \instate\,.
\end{equation}
This is just the tree-level subleading soft theorem \eqref{tree level subleading theorem}.

The equivalence with the conservation of superrotation charges then proceeds as before. The hard part of flux is used to generate the superrotation transformation of the hard external wavefunctions,
\begin{equation}
F_{\mY}^{hard} |\text{out}\rangle=-i \delta_{\mY} |\text{out}\rangle\,,
\end{equation}
while the soft part of the flux inserts the field $\mNone_{\bz \bz}$ into the $\mathcal{S}$-matrix elements. Importantly, the soft flux used in \cite{Kapec:2014opa} only contained the first term in \eqref{FY soft}, namely
\begin{equation}
F_{\mY}^{soft,0}=-\frac{2}{\kappa^2} \int_{\mathcal{S}} d^2z\, \mY\, \partial^3 \mathcal{N}^{(1)}_{\bz \bz}\,.
\end{equation}
Here and in what follows, we set up our computations in a superrotation frame where the superboost fields vanish $\varphi(z)=\bar \varphi(\bz)=0$ without loss of generality, such that the conformal covariant derivative becomes a simple derivative $\mathscr{D} \mapsto \partial$. 
With this expression of the soft flux, conservation of the superrotation charges is found equivalent to the tree-level subleading soft theorem \eqref{N1 insertion}.

While the leading soft factor $\hat S^{(0)}_n$ is tree-level exact, the subleading soft factor $S^{(1)}_n$ receives corrections which are one-loop exact \cite{Bern:2014oka,He:2014bga}.\footnote{It was argued in \cite{Cachazo:2014dia}
 that a different approach consisting of taking the soft limit \emph{before}
the expansion in the regularisation parameter $\epsilon$ can lead to soft factors which are uncorrected at loop orders. The way how to make
this new prescription fully consistent with the requirement that divergences cancel in
cross-sections is still an open question \cite{Bern:2014oka}.} For \textit{massless} external particles, the perturbatively exact soft theorem can be written up to subleading order \cite{Bern:2014oka,He:2017fsb},
\begin{equation}
\label{exact soft theorem}
\mathcal{M}_{n+1} =\left[\omega^{-1}\, \hat S^{(0)}_n+S^{(1)}_n+ \kappa^2 \left( \epsilon^{-1} \hat \sigma_{n+1}' \hat S^{(0)}_n  - \epsilon^{-1} \left(S^{(1)}_n \sigma_n\right)+ \Delta S_n^{\text{fin}} \right)\right]\mathcal{M}_n+O(\omega)\,,
\end{equation}
with
\begin{equation}
\label{sigma_n}
\sigma_n =\frac{\hbar}{(8\pi)^2} \sum_{i,j=1}^n (p_i \cdot p_j) \ln \frac{p_i \cdot p_j}{\mu}\,.
\end{equation}
The terms in parenthesis multiplying the factor $\kappa^2$ in \eqref{exact soft theorem} are the one-loop exact corrections to the subleading soft factor. Apart from the \textit{potential} finite contribution $\Delta S^{\text{fin}}_n$, these one-loop corrections are infrared divergent in the limit $\epsilon \to 0$. In specific examples no finite contributions $\Delta S^{\text{fin}}_n$ has been found \cite{Bern:2014oka,He:2014bga}, although a general argument that it should always vanish has not been provided thus far. We will comment on this later on. Also note that the operator $S_n^{(1)}$ only acts on the scalar $\sigma_n$. 

A few years ago He, Kapec, Raclariu and Strominger proposed that the one-loop correction to the subleading soft factor can be accounted for by the addition of new terms in the soft part of superrotation flux \cite{He:2017fsb}, 
\begin{equation}
\label{HKRS soft flux}
F^{soft,\text{HKRS}}_{\mY}=F^{soft,0}_{\mY}- \frac{i\hbar}{\epsilon} \frac{1}{\pi \kappa^2} \int_{\mathcal{S}} d^2z\, \mY \left[2\, \mNzero_{zz} \partial \mNzero_{\bz \bz} + \partial ( \mNzero_{zz} \mNzero_{\bz \bz})\right]\,.
\end{equation}
Indeed using the leading soft theorem in the form of \eqref{N0 insertion}, insertion of the soft flux \eqref{HKRS soft flux} into the amplitude $\outstate \mathcal{S} \instate$ reproduces the IR-divergent one-loop corrections to the subleading soft theorem \eqref{exact soft theorem}. Here it should be noted that the additional terms in the soft flux \eqref{HKRS soft flux} explictly depend on the parameters $\hbar$ and $\epsilon$. Hence its interpretation would be a one-loop renormalisation of the soft flux containing explicit IR divergences. 

Instead, we show that the one-loop corrections to the subleading soft factor are already taken care of by the \textit{classical} soft flux \eqref{FY soft}. In particular, we show that they arise through insertion of the terms missed in previous literature, namely
\begin{equation}
\label{new soft terms}
F_{\mY}^{soft,new}=\frac{2}{\kappa^2}\int_{\mathcal{S}} d^2 z\,  \mathcal{Y}  \left[-\partial^3 ( C^{(0)} \mathcal{N}^{(0)}_{\bar z \bar z} ) +3 \bar \partial^2\mathcal{N}^{(0)}_{zz} \partial C^{(0)}+ C^{(0)} \partial\bar\partial^2 \mathcal{N}_{zz}^{(0)}  \right]\,.
\end{equation}
Note that these terms do not explicitly depend on $\hbar$ nor $\epsilon$. The way these factors eventually appear in the subleading soft factor can already be anticipated: they naturally arise from insertion of the supertranslation Goldstone mode $C^{(0)}$ through \eqref{C0 insertion}! 

Therefore, we want to demonstrate that insertion of \eqref{new soft terms} reproduces the (integrated) one-loop corrections to \eqref{exact soft theorem}, namely 
\begin{align}
\label{goal}
\outstate F_{\mY}^{soft,new} \mathcal{S}\instate=- \frac{i\kappa}{8\pi\epsilon}  \int_{\mathcal{S}} d^2z\,  \mY\, \partial^3 \left(\hat \sigma_{n+1}' \hat S^{(0)-}_n  - \left(S^{(1)-}_n \sigma_n \right) \right) \outstate \mathcal{S} \instate\,.
\end{align}
Following the computations in \cite{He:2017fsb}, we can write
\begin{align}
S^{(1)-}_n \sigma_n &=\frac{\kappa \hbar}{(8\pi)^2} \sum_{i,j=1}^n \left[\frac{p_j \cdot \hat q}{p_i \cdot \hat q}\, p_i^\mu p_i^\nu -p_i^\mu p_j^\mu \right] \varepsilon^-_{\mu\nu}(\hat q)\, \ln \frac{p_i \cdot p_j}{\mu}\,,
\end{align}
such that
\begin{equation}
    \begin{split}
\label{second term}
&\partial^3 \left(S^{(1)-}_n \sigma_n \right)\\
&=-\frac{\kappa \hbar}{32\pi} \sum_{i,j=1}^n \omega_i \left[ (p_j \cdot \hat q) \partial \delta^{(2)}(z-z_i)+3 (p_j \cdot \partial \hat q) \delta^{(2)}(z-z_i)\right] \ln \frac{p_i \cdot p_j}{\mu}\\
&=-\frac{\kappa \hbar}{32\pi} \sum_{i,j=1}^n \omega_i \left[ (p_j \cdot \hat q) \partial \delta^{(2)}(z-z_i)+3 (p_j \cdot \partial \hat q) \delta^{(2)}(z-z_i)\right] \ln \frac{\hat q \cdot p_j}{\mu}\\
&=\hat{\sigma}'_{n+1}\, \partial \bar\partial^2 \hat S^{(0)+}_n+3\, \partial \hat{\sigma}'_{n+1}\, \bar \partial^2 \hat S^{(0)+}_n \,.
\end{split}
\end{equation}
In going from the first to the second line, we have used $\partial^2 \hat q^\mu=0$ together with
\begin{equation}
\label{D^2 S0}
\partial^2 \left( \frac{p_i^\mu p_i^\nu \varepsilon_{\mu\nu}^-(\hat q)}{p_i \cdot \hat q} \right)=-2\pi \omega_i\,  \delta^{(2)}(z-z_i)\,.
\end{equation}
To get to the third line we have used momentum conservation $\sum_j p_j^\mu=0$.\footnote{Note that we can alternatively use the momentum conservation $\sum_j p_j^\mu+\omega \hat{q}^\mu=0$ associated with the scattering $\mathcal{M}_{n+1}$. This does not affect the computations in this paper due to the properties $\varepsilon^\pm \cdot \hat{q}=\hat{q} \cdot \hat{q}=0$.} To obtain the final expression, we used the definitions \eqref{S0} and \eqref{sigma_n} together with \eqref{D^2 S0}. Therefore the one-loop correction \eqref{goal} can be alternatively written
\begin{align}
\nonumber
&\outstate F_{\mY}^{soft,new} \mathcal{S}\instate\\
\label{goal bis}
&= \frac{i\kappa}{8\pi\epsilon}  \int_{\mathcal{S}} d^2z\,  \mY \left[- \partial^3(\hat \sigma_{n+1}' \hat S^{(0)-}_n)  +\hat{\sigma}'_{n+1}\, \partial \bar \partial^2 \hat S^{(0)+}_n+3\, \partial \hat{\sigma}'_{n+1}\, \bar \partial^2 \hat S^{(0)+}_n  \right] \outstate \mathcal{S} \instate\,.
\end{align}
In this form, it is straightforward to see that it indeed coincides with the insertion of \eqref{new soft terms} upon using \eqref{leading soft theorem} and \eqref{C0 insertion}.

Our results are consistent with vanishing finite correction $\Delta S_n^{\text{fin}}=0$. However, this conclusion appears to depend on the renormalisation scheme. In promoting the soft flux \eqref{new soft terms} to a quantum operator we had to remove the UV divergence associated with the composite operator $C^{(0)}(z,\bz)\, \mNzero_{zz}(z,\bz)$. We did this by point-splitting regularisation, i.e., by simply discarding the Wick contraction of $C^{(0)}$ and $\mNzero_{zz}$ when evaluated at the same point. This latter contribution, although UV-divergent, is IR-finite and a different regularisation scheme could potentially give a nonzero contribution to $\Delta S_n^{\text{fin}}$. 

The one-loop correction to the subleading soft graviton theorem in the form \eqref{goal bis} equally applies to the case of massive external particles, since it straightforwardly follows from insertion of the soft flux contribution \eqref{new soft terms} into the $\mathcal{S}$-matrix elements. However it cannot be written in the form \eqref{goal} since the identities \eqref{second term} and \eqref{D^2 S0} only hold for massless momenta. We can massage \eqref{goal bis} in order to find a universal expression for the one-loop correction to the subleading soft factor. Using the identity given in \eqref{D^2 sigma'} and equally valid for massless and massive momenta,
\begin{equation}
\partial^2 \hat \sigma_{n+1}' =\frac{\hbar}{8\pi^2 \kappa}\, \hat S^{(0)+}_n\,,
\end{equation}
and integrating by parts, we have
\begin{align}
&\int_{\mathcal{S}} d^2z\,  \mY \left[\hat{\sigma}'_{n+1} \partial \bar \partial^2 \hat S^{(0)+}_n+3 \partial \hat{\sigma}'_{n+1} \bar \partial^2 \hat S^{(0)+}_n  \right]\nonumber\\
&=\frac{1}{2}\int_{\mathcal{S}} d^2z\,  \mY \left[\hat{\sigma}'_{n+1} \partial \bar \partial^2 \hat S^{(0)+}_n+3 \partial \hat{\sigma}'_{n+1} \bar \partial^2 \hat S^{(0)+}_n +\bar \partial^2 \hat{\sigma}'_{n+1} \partial  \hat S^{(0)+}_n+3 \partial \bar \partial^2 \hat{\sigma}'_{n+1} \hat S^{(0)+}_n  \right]\nonumber\\
&=\frac{4\pi^2\kappa}{\hbar} \int_{\mathcal{S}} d^2z\, \mY\, \partial^3 \left(\hat{\sigma}'_{n+1}\, \bar \partial^2 \hat{\sigma}'_{n+1} \right)=\frac{1}{2}\int_{\mathcal{S}} d^2z\, \mY\, \partial^3 (\hat{\sigma}'_{n+1}\, \hat S^{(0)-}_n )\,.
\end{align}
This allows the drastic simplification
\begin{align}
\outstate F_{\mY}^{soft,new} \mathcal{S}\instate=- \frac{i\kappa}{16\pi\epsilon}  \int_{\mathcal{S}} d^2z\,  \mY\, \partial^3(\hat \sigma_{n+1}' \hat S^{(0)-}_n)\,   \outstate \mathcal{S} \instate\,,
\end{align}
from which we deduce a formula for the one-loop correction to the subleading soft graviton theorem valid whether external momenta are massive or massless,
\begin{equation}
\mathcal{M}_{n+1}\big|_{\text{1-loop}} =\frac{\kappa^2}{2\epsilon} \left( \hat \sigma_{n+1}'\, \hat S^{(0)}_n+\ker(\partial^3) \right) \mathcal{M}_n+O(\omega)\,,
\end{equation}
with the caveat that we are missing terms annihilated by the operator $\partial^3$. In view of this new result, it would be interesting to revisit the one-loop correction to the subleading soft factor when massive particles are considered.

\section{The celestial stress tensor}
\label{sec:Celestial stress-tensor}
The results of the previous sections have additional important implications from the perspective of celestial holography, as already anticipated and discussed in \cite{He:2017fsb}. Indeed they confirm that the celestial stress tensor of four-dimensional \textit{quantum gravity} is the soft superrotation flux transforming appropriately in the coadjoint representation of the extended BMS group as proposed recently in \cite{Donnay:2021wrk}. Using the parametrisation of the flux given in \eqref{superrotation fluxes} and making the choice $\mY(w)=\frac{1}{z-w}$ for the symmetry parameter, the celestial stress tensor thus reads
\begin{equation}
\label{celestial T}
T(z)= \frac{i}{8\pi G}\int_\mathcal{S} d^2 w\,  \frac{1}{z-w}  \left(-{\mathscr D}^3 \mathcal N_{\bar w \bar w}^{(1)} + \frac{3}{2} C^{(0)}_{ww} \mathscr{D} \mathcal{N}^{(0)}_{\bar w \bar w}+ \frac{1}{2}\mathcal{N}^{(0)}_{\bar w \bar w} \mathscr{D} C_{ww}^{(0)}  \right)\,.
\end{equation}
This extends the expression given in \cite{Kapec:2016jld} beyond tree-level and corrects the one originally suggested in \cite{He:2017fsb}. As an immediate consequence, we also conclude that the associated celestial central charge strictly vanishes \textit{at all orders in perturbation theory} since the soft flux does not transform anomalously.  

The subleading soft graviton theorem is in direct correspondence with the conformal Ward identity of the celestial stress tensor \eqref{celestial T}. To make this completely manifest, one needs to turn to a basis of \textit{boost} eigenstates rather than the more standard momentum basis. This can be achieved by Mellin transform of the operators $\mO_\omega(z,\bz)$ associated with hard external particle insertions \cite{Pasterski:2016qvg,Pasterski:2017kqt,Pasterski:2017ylz},
\begin{equation}
\mO_{h,\bar h}(z,\bz)= \int_0^\infty d\omega\, \omega^{\Delta-1}\, \mO_\omega(z,\bz)\,,
\end{equation}
where $(h,\bar h)$ are related to the conformal dimension and spin of the operator through $\Delta=h+\bar h$ and $J=h-\bar h$. In this basis the subleading soft graviton theorem \eqref{exact soft theorem} takes the form \cite{Kapec:2016jld,He:2017fsb} 
\begin{equation}
\label{Ward identity}
_{boost}\outstate T(z) \mathcal{S} \instate_{boost}=\sum_{i=1}^n\left[\frac{h_i}{(z-z_i)^2}+\frac{\partial_{z_i}}{z-z_i} \right] {}_{boost}\outstate \mathcal{S} \instate_{boost}\,,
\end{equation}
which is just the conformal Ward identity of a two-dimensional celestial CFT, with celestial amplitudes given by
\begin{equation}
\langle \prod_{i=1}^n \mO_{h_i,\bar h_i}(z_i,\bz_i) \rangle_{\text{CCFT}} \equiv {}_{boost}\outstate \mathcal{S} \instate_{boost}\,.
\end{equation}
Because the subleading soft theorem \eqref{exact soft theorem} and the conformal Ward identity \eqref{Ward identity} are one loop-exact, so is the celestial stress tensor \eqref{celestial T}. This offers exciting perspectives for a celestial holographic description of quantum gravity in asymptotically flat spacetimes.

\section*{Acknowledgments}

We thank Adrien Fiorucci, Laurent Freidel, Alok Laddha and Andrea Puhm for illuminating discussions. We especially thank Sabrina Pasterski for pointing out a possible connection between quadratic terms in the soft flux and one-loop corrections to the subleading soft graviton theorem.  We are also grateful to Francesco Alessio, Mina Himwich, Ali Seraj for useful correspondence. LD and RR are supported by the Austrian Science Fund (FWF) START project Y 1447-N. The work of KN is supported by the ERC Consolidator Grant N.~681908 “Quantum black holes: A microscopic window into the microstructure of gravity” and by the STFC grants ST/P000258/1 and ST/T000759/1.

\appendix

\section{Newman-Penrose quantities}
\label{sec:Newman-Penrose formalism vs Metric formalism}

In this appendix, we provide some useful formulas in the Newman-Penrose formalism \cite{Newman:1961qr, Newman:1962cia}, as well as the dictionary with the metric formulation. We follow the conventions of \cite{Barnich:2011ty, Barnich:2013axa, Barnich:2019vzx}.

The asymptotic Weyl scalars satisfy 
\begin{equation}
    \begin{split}
        &\Psi^0_4 = - \Ddot{\bar{\sigma}}^0 \,, \qquad \Psi^0_3 = - \bar{\partial} \dot{\bar{\sigma}}^0 \,, \\
        &\Psi^0_2 - \bar{\Psi}^0_2 = \bar{\eth}^2 \sigma^0 - \eth^2 \bar{\sigma}^0 + \dot{\sigma}^0 \bar{\sigma}^0 - \sigma^0 \dot{\bar{\sigma}}^0\,,
        \end{split}
\end{equation} together with the time evolution equations
\begin{equation}
    \begin{split}
        &\partial_u \Psi^0_3 = \bar{\partial} \Psi^0_4 \,, \\
        &\partial_u \Psi^0_2 =  \bar{\partial} \Psi^0_3 + \sigma^0 \Psi^0_4 \,, \\
        &\partial_u \Psi^0_1 =  \bar{\partial} \Psi^0_2 + 2 \sigma^0 \Psi^0_3 \,.
    \end{split}
\end{equation} The relation between the Newman-Penrose scalars and the functions appearing in the expansion of the metric \eqref{Bondi gauge metric} is given by \cite{Barnich:2011ty, Freidel:2021qpz} 
\begin{equation}
\begin{split}
    &\sigma^0 = \frac{1}{2} C_{\bar{z}\bar{z}} \,, \qquad \dot{\sigma}^0 = \frac{1}{2} N_{\bar{z}\bar{z}} \,,\\
    &\Psi^0_2 + \bar{\Psi}^0_2 = -2 M - \frac{1}{4} (N_{\bar{z}\bar{z}} C^{\bar{z}\bar{z}} + N_{zz} C^{zz}) \,, \\
    &\Psi^0_2 - \bar{\Psi}^0_2 = \frac{1}{2} \partial^2  C^{zz}  - \frac{1}{2}  \bar{\partial}^2 C^{\bar{z}\bar{z}} + \frac{1}{4}N_{\bar{z}\bar{z}} C^{\bar{z}\bar{z}} - \frac{1}{4} N_{zz} C^{zz} \,, \\
    &\Psi^0_1 = - \left[ N_{\bar{z}} + \frac{1}{4}C_{\bar{z}\bar{z}} \bar{\partial} C^{\bar{z}\bar{z}} + \frac{3}{16} \bar{\partial}  (C_{zz} C^{zz}) \right] \,.
\end{split}
\end{equation}

\section{Massive external particles}
\label{app:massive}
In this appendix we provide details regarding the case of massive external particles. We parametrise massive momenta by
\begin{equation}
p^\mu=\frac{m}{2\rho} \left(1+\rho^2(1+ z\bz)\,, \rho^2 (z+\bz)\,, -i \rho^2(z-\bz)\,,-1+\rho^2(1-z\bz)  \right)\equiv m\, \hat p^\mu\,,
\end{equation}
where $(\rho,z,\bz)$ refer to a point of the three-dimensional hyperboloid describing the direction of approach of the particle to $i^+$. More specifically, it relies on the hyperbolic slicing of Minkowski space given by
\begin{equation}
ds^2=-d\tau^2+\tau^2 \left(\frac{d\rho^2}{\rho^2}+\rho^2 dz d\bz \right)\,.
\end{equation}
For massive particles, the vertex operators are given by \cite{Himwich:2020rro}
\begin{equation}
\mW_m(p^\mu)=\exp\left[\frac{im}{2}\int_\mathcal{S} d^2w\, G(\hat p;w,\bar w)\, C^{(0)}(w,\bar w)\right]\,,
\end{equation}
where $G(\hat p;w,\bar w)$ is the bulk-boundary propagator \cite{Campiglia:2015lxa}
\begin{equation}
G(\hat p;w,\bar w)=\frac{1}{\pi} \left(\frac{\rho}{1+\rho^2|z-w|^2} \right)^3\,,
\end{equation}
satisfying 
\begin{equation}
\lim_{\rho \to \infty} \rho^{-1} G(\hat p;w,\bar w)=\delta^{(2)}(z-w)\,.
\end{equation}
Their correlation functions are given by \cite{Himwich:2020rro}
\begin{equation}
\langle \mW_1\, ...\, \mW_n \rangle=\exp \left[-\frac{1}{8} \sum_{i,j} m_i m_j \int_\mathcal{S} d^2w\, d^2z\, G(\hat p_i;w) G(\hat p_j;z) \langle C^{(0)}(w) C^{(0)}(z) \rangle \right]\,.
\end{equation}
This allows us to compute the effect of inserting the supertranslation Goldstone mode $C^{(0)}$ into the $\mathcal{S}$-matrix elements,
\begin{subequations}
\begin{align}
\langle C^{(0)}(z) \mW_1\, ...\, \mW_n \rangle&=-2i \lim_{\rho \to \infty} \rho^{-1}\, \partial_m \langle \mW_m(\rho,z,\bz) \mW_1\, ...\, \mW_n \rangle\big|_{m=0}\\
&=\frac{i}{2} \sum_{i=1}^n m_i \int_\mathcal{S} d^2w\, G(\hat p_i;w) \langle C^{(0)}(z) C^{(0)}(w) \rangle\, \langle \mW_1\, ...\, \mW_n \rangle\,.
\end{align}
\end{subequations}
The integral appearing in the above equation can be computed using the fact that $\langle C^{(0)} C^{(0)} \rangle$ as given in \eqref{CC correlator} is the Green's function of the operator $\partial^2 \bar \partial^2$, 
\begin{align}
\bar \partial^2 \partial^2 \int_\mathcal{S} d^2w\, G(\hat p_i;w) \langle C^{(0)}(z) C^{(0)}(w) \rangle=\frac{\hbar \kappa^2}{8\pi \epsilon}\, G(\hat p_i;z) =-\frac{\hbar \kappa^2}{8\pi^2 m_i\, \epsilon}\, \bar \partial^2\left( \frac{\varepsilon_{\mu\nu}^+ p_i^\mu p_i^\nu}{p_i \cdot \hat q}\right)\,,
\end{align}
where the last equality is one of the important properties of the bulk-boundary propagator $G(\hat p;z)$ \cite{Campiglia:2015kxa,Campiglia:2015lxa}. Thus we have
\begin{align}
\partial^2 \langle C^{(0)}(z,\bar{z}) \mW_1\, ...\, \mW_n \rangle&=-\frac{i\hbar \kappa}{8\pi^2 \epsilon}\, \hat S^{(0)+}_n  \langle \mW_1\, ...\, \mW_n \rangle\,.
\end{align}
In order to integrate this equality we use the following relation,
\begin{align}
\label{D^2 sigma'}
\partial^2 \hat \sigma_{n+1}'=\frac{\hbar}{2(4\pi)^2} \sum_i \frac{( p_i \cdot \partial \hat q)^2}{ p_i \cdot \hat q}=\frac{\hbar}{(4\pi)^2} \sum_i \frac{( p_i \cdot \varepsilon^+)^2}{ p_i \cdot \hat q} =\frac{\hbar}{8\pi^2 \kappa}\, \hat S^{(0)+}_n\,,
\end{align}
such that we can write
\begin{equation}
\langle C^{(0)}(z,\bar{z}) \mW_1\, ...\, \mW_n \rangle = -\frac{i\kappa^2}{\epsilon}\, \hat \sigma'_{n+1}\,  \langle \mW_1\, ...\, \mW_n \rangle\,.
\end{equation}
This formula is identical to \eqref{C0 insertion} for the insertion of $C^{(0)}$ when external particles are massless.

\hypersetup{urlcolor=blue}
\bibliographystyle{utphys}
\bibliography{references}
\end{document}